# Mapeo sistemático sobre estudios empíricos realizados con colecciones de proyectos software

Juan Andrés Carruthers[1], Jorge Andrés Diaz Pace[2], Emanuel Agustín Irrazábal[1]

[1] Universidad Nacional del Nordeste, Departamento de Informática, Corrientes, Argentina

[2] Universidad Nacional del Centro de la Provincia de Buenos Aires, Instituto Superior De Ingeniería Del Software, Buenos Aires, Argentina

{jacarruthers, eairrazabal}@exa.unne.edu.ar

adiaz@exa.unicen.edu.ar

**Resumen.** Contexto: los proyectos software son insumos comunes en los experimentos de la Ingeniería del Software, aunque estos muchas veces sean seleccionados sin seguir una estrategia específica, lo cual disminuye la representatividad y replicación de los resultados. Una opción es el uso de colecciones preservadas de proyectos software, pero estas deben ser vigentes y con reglas explícitas que aseguren su actualización a lo largo del tiempo. Objetivo: realizar un estudio secundario sistematizado sobre las estrategias de selección de los proyectos software en estudios empíricos para conocer las reglas tenidas en cuenta, el grado de uso de colecciones de proyectos, los meta-datos extraídos y los análisis estadísticos posteriores realizados. Método: se utilizó un mapeo sistemático para identificar estudios publicados desde enero de 2013 a diciembre de 2020. Resultados: se identificaron 122 estudios de los cuales el 72% utilizó reglas propias para la selección de proyectos y un 27% usó colecciones de proyectos existentes. Asimismo, no se encontraron evidencias de un marco estandarizado para la selección de proyectos, ni la aplicación de métodos estadísticos que se relacionen con la estrategia de recolección de las muestras.

**Palabras Clave.** Colecciones, Proyectos software, Experimentación, Ingeniería del Software Basada en Evidencia.

## 1. Introducción

El desarrollo software actual trabaja con la construcción de aplicaciones multi versión [1] que crecen, tanto en complejidad como en funcionalidad, siendo necesario conservar su calidad actual [2]. Por ello, es necesario obtener métodos empíricos para demostrar la calidad del producto software [3] y utilizar evidencia directamente relacionada con el producto software resultante a partir de mediciones que se vinculen con los atributos de calidad del código fuente [4].

En este sentido, la Ingeniería del Software Basada en Evidencia (ISBE) proporciona los medios para que la evidencia actual de la investigación pueda integrarse con la experiencia práctica y los valores humanos en el proceso de toma de decisiones para el desarrollo y mantenimiento de software [5]. En el caso de los experimentos cuyo objeto de estudio es el código fuente, una colección ad hoc de proyectos software muchas veces no es suficiente para lograr representatividad o replicación en los resultados. Con la intención de mejorar los resultados surgen las colecciones preservadas de proyectos software, que se utilizan para reducir el costo de recopilar los proyectos software y para que los estudios sean replicables y comparables [6].

Estas colecciones son un insumo para los grupos de trabajo y sirven como mecanismo de comparación para los experimentos en la disciplina de Ingeniería del Software. Existen distintas colecciones, como las realizadas por Barone y Sennrich [7], Allamanis y Sutton [8] o Keivanloo [9], que se diferencian por la cantidad, calidad de proyectos que lo componen o los criterios y métodos de agrupamiento. Por ejemplo, Zerouali y Mens [10] investigaron el uso de las bibliotecas y frameworks de pruebas en Java mientras que Goeminne y Mens [11] trabajaron con frameworks de bases de datos.

En este y otros ejemplos las reglas para la selección de los proyectos muchas veces no son explícitas, se corresponden con selecciones anteriores o son al azar. La definición de un modelo de procedimientos a partir de reglas estándar y herramientas es fundamental para garantizar la capacidad de preservar sistemáticamente la colección a lo largo del tiempo por integrantes del mismo equipo de trabajo o externos, es decir, que la conservación de la misma sea independiente del grupo que la construyó. Así, por ejemplo,

la última versión del Qualitas es del año 2013 [6], lo que hace necesaria la revisión de los proyectos y sus versiones y la generación de los meta-datos necesarios. En otras colecciones, con mayor cantidad de proyectos [8], su diseño no ha tenido en cuenta la extracción de meta-datos necesarios para el estudio de la calidad del producto software.

Por lo antes indicado, el objetivo de este trabajo es determinar los criterios de selección de los proyectos software que son insumos de estudios empíricos, las características de estos proyectos, qué meta-datos se extraen, qué herramientas se utilizan para obtener los meta-datos y qué análisis estadísticos se realizan con los meta-datos recopilados. Se eligió el enfoque de mapeo sistemático para obtener una visión general del desarrollo técnico actual o el nivel de práctica de un área de investigación [12]. Este estudio pretende brindar una visión general sobre las prácticas seguidas por los grupos de investigación para la experimentación con colecciones de proyecto, exponiendo los problemas encontrados que comprometen a la representatividad de las muestras, la replicación de los experimentos y la generación de los resultados, y de esta manera evitar que siga sucediendo en futuros trabajos.

Además de esta introducción, el trabajo se encuentra organizado de la siguiente manera. La sección 2 describe los estudios relacionados con la temática. En la sección 3 se detalla la metodología empleada para el estudio y se presentan las preguntas de investigación. En la sección 4 se reportan las actividades llevadas a cabo. En la sección 5 se discuten y se responden las preguntas de investigación. Finalmente, en la sección 6 se incluye la conclusión del mapeo sistemático desarrollado.

## 2. Trabajos relacionados

El uso de proyectos software para realización de estudios empíricos no resulta una práctica novedosa en la Ingeniería de Software. En 1971, Knuth publicó su artículo [13] en el cual recolectó aleatoriamente programas en FORTRAN de la Universidad de Stanford y de la compañía Lockheed Missiles and Space. Chidamber y Kemerer en [14] desarrollaron un trabajo para validar las métricas orientadas a objetos creadas por ellos mismos tres años antes [15], una parte del mismo buscaba demostrar la factibilidad de estas métricas en dos sistemas comerciales: uno codificado en el lenguaje C++ y el otro en Smalltalk. Harrison et al. [16] también condujeron un estudio con cinco proyectos software en tecnología C++ para evaluar y comparar la métrica acoplamiento entre objetos y la métrica número de asociaciones entre una clase y sus pares.

El advenimiento de plataformas para compartir código como SourceForge, aumentó la accesibilidad a proyectos de fuente abierta [6]. Con estos nuevos repositorios públicos fue posible el acceso al código fuente versionado de proyectos software con distintos tamaños, tecnología o perfiles de trabajo de los equipos.

Esto hizo más necesaria la construcción de colecciones de proyectos software que aumente la replicabilidad de los experimentos, se agreguen los resultados, se reduzcan los costos y se mejore la representatividad de las muestras. En la literatura existe una gran variedad de ejemplos de estas colecciones. Barone y Sennrich [7] crearon una colección de más de 100K funciones en Python con sus respectivos cuerpos y descripciones para estudios.

Allamanis y Sutton [8] construyeron el Github Java Corpus con todos los proyectos Java en Github Archive que no fuesen repositorios duplicados. Similar al anterior, Keivanloo et al. [9] incluyeron 24824 proyectos Java, alojados en SourceForge y Google Code.

Zerouali y Mens [10] actualizaron el Github Java Corpus y analizaron el uso de bibliotecas y frameworks de pruebas como JUnit, Spring o TestNG. En esta actualización agregaron los repositorios creados en Github no presentes en la colección original y descartaron aquellos que no estuvieran disponibles y los que no utilizaron la herramienta Maven para gestionar las dependencias, teniendo como resultado una colección de 4532 proyectos. De igual manera, Goeminne y Mens [11] realizaron cambios al Github Java Corpus y estudiaron cinco frameworks de base de datos.

Tempero et al. [6] en el año 2013 constituyeron la colección Qualitas Corpus donde incluyeron sistemas desarrollados en Java con sus archivos fuente y binarios disponibles públicamente en formato ".jar". El objetivo del Qualitas es reducir sustancialmente el costo de los equipos de investigación para desarrollar grandes estudios empíricos del código fuente.

En algunas colecciones, además de la documentación, código fuente y binarios del proyecto, también se proporcionan meta-datos relacionados. Por ejemplo, FLOSSmole [17] brinda los nombres de los proyectos, lenguajes de programación, plataformas, tipo de licencia, sistema operativo y datos de los desarrolladores involucrados. FLOSSMetrics [18] calcula, extrae y almacena información de sistemas de control de versiones, sistemas de rastreo de defectos, archivos de listas de correos y métricas de código. Qualitas.class [19] por su parte, contiene medidas relativas a los proyectos tales como métricas de tamaño (cantidad de líneas de código, número de paquetes, clases e interfaces) y métricas de diseño.

Finalmente, mapeos sistemáticos como los de Falessi et al. [20] y Cosentino et al. [21] estudiaron los conjuntos de datos de proyectos usados en la Ingeniería del Software, debido a la gran diversidad de criterios considerados para seleccionar los proyectos y los meta-datos obtenidos. En ambos trabajos reportaron un bajo nivel replicabilidad de las metodologías empleadas para su recolección.

## 3. Metodología

Se llevó a cabo un estudio de mapeo sistemático siguiendo las pautas identificadas en [22] para obtener una visión general del uso de colecciones de proyectos en la ISBE. Se seleccionó esta técnica por centrarse en la "clasificación y análisis temático de un tema de la Ingeniería del Software" [23]. En este caso, aunque la recolección de proyectos sea una práctica estándar para la experimentación en ISBE, los criterios de selección de los proyectos software, sus características, los meta-datos que se extraen de estos proyectos, qué herramientas se utilizan para obtener los meta-datos, así como también qué análisis estadísticos se realizan con los meta-datos recopilados no han sido abordados ampliamente por otros estudios secundarios. Por lo tanto, es necesario identificar, categorizar y analizar la investigación disponible sobre el tema para describir las prácticas y obtener una visión general de su estado de arte.

Las preguntas de investigación que guiaron el desarrollo del estudio se presentan en la Tabla I.

**Tabla I:** Preguntas de investigación.

| Pregunta de investigación | Motivación |
|---|---|
| **RQ1**: ¿Cuáles son los criterios de selección de los proyectos software objeto de estudios empíricos? | Identificar cuáles son los criterios tenidos en cuenta por los investigadores para la selección de proyectos en la realización de estudios empíricos. |
| **RQ2**: ¿Con qué tipo de proyectos trabajan los grupos de investigación para realizar estudios empíricos? | Conocer qué características tienen los proyectos seleccionados en términos del tipo de software y lenguaje de programación. |
| **RQ3**: ¿Cuáles son los datos/meta-datos que se extraen de estos proyectos? | Determinar los meta-datos y métricas que son extraídos de cada uno de estos proyectos para su posterior experimentación. |
| **RQ4**: ¿Qué herramientas se utilizan para obtener estos datos/meta-datos? | Determinar cuáles son las herramientas usadas en los diferentes estudios para recolectar los meta-datos. |
| **RQ5**: ¿Qué análisis estadísticos se realizan con los datos/meta-datos recopilados? | Definir los tipos de análisis estadísticos que se desarrollan sobre los experimentos hechos con los proyectos y sus respectivos meta-datos para interpretar los resultados obtenidos. |

### 3.1. Selección de artículos

Para reunir los estudios primarios relevantes se llevó a cabo una búsqueda y selección iterativa en tres fases tal y como se indica en la Fig. I.

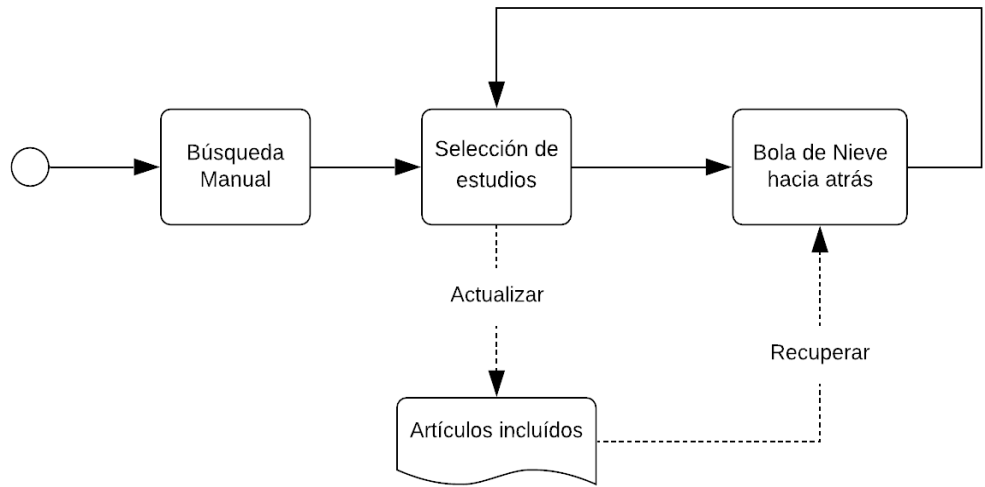

Fig. I: Proceso de búsqueda y selección de artículos.

**Búsqueda manual**: se realizó una búsqueda manual en las principales conferencias y revistas enfocadas en ISBE. Se seleccionó la revista Empirical Software Engineering (EMSE) y las conferencias Empirical Software Engineering and Measurement (ESEM) e International Conference on Evaluation and Assessment in Software Engineering (EASE) por ser representativas del área de investigación y haber sido utilizadas en otros estudios, como en [24] y en [25].

De acuerdo a estrategias planteadas en trabajos como [24], [26] y las buenas prácticas de búsquedas manuales de [22], en primera instancia se seleccionó el período de tiempo comprendido entre el 1 de enero del 2013 hasta el 30 de noviembre del 2018 de ESE y ESEM, de forma análoga a [24]. Posteriormente se complementó el rango de búsqueda hasta el 31 de diciembre del 2020 conforme se fueron publicando nuevos números de la revista y el congreso. Después se incorporaron las ediciones del congreso EASE como en [25] considerando el mismo período de tiempo. De esta manera se tomaron las publicaciones de EMSE, ESEM y EASE de los últimos 8 años.

En particular, el instrumento de recolección fue una hoja de cálculo, donde se escribieron los nombres y autores de los artículos de la revista y las conferencias.

**Selección de estudios:** esta fase consistió en una revisión dual que se realiza de forma iterativa. Los artículos candidatos, después de ser recopilados se incluyeron o excluyeron de acuerdo con los criterios presentes en la Tabla II. Los trabajos fueron analizados considerando resumen, introducción, metodología, resultados y conclusión. En cada iteración, se seleccionaron 15 estudios al azar que fueron revisados por dos investigadores. Los mismos anotaron sus decisiones de incluir o excluir cada estudio, junto con el CI o CE en que se basó la decisión. Para medir el grado de acuerdo entre los investigadores se utilizó el estadístico Kappa de Cohen, como sugieren [27] y [28]. El valor registrado de Kappa de Cohen fue 0.77, lo que demuestra un nivel de acuerdo alto.

**Muestreo "bola de nieve":** se complementó la búsqueda manual con el método de bola de nieve hacia atrás con un muestreo de los artículos incluidos. Las referencias proporcionaron artículos relacionados o similares. Aquellos artículos recopilados durante esta etapa también se agregaron a la lista de candidatos y se seleccionaron de acuerdo con los criterios descritos anteriormente. Este proceso se realizó en dos iteraciones.

Tabla II: Criterios de inclusión y exclusión de estudios.

| Identificador | Descripción |
| --- | --- |
| CI1 | El artículo pertenece a ESE, EASE y ESEM. |
| CI2 | Es un artículo completo. |
| CI3 | En el artículo se realizan estudios empíricos. |
| CI4 | Los estudios empíricos se realizan en base a la selección de un conjunto de proyectos software. |
| CE1 | Artículos no técnicos (guías, artículos de introducción, editoriales, etc.) |
| CE2 | Artículos duplicados. |

**Evaluación de la calidad:** en este trabajo se decidió no realizar una evaluación de calidad de los estudios primarios, al contrario de lo que se propone en las guías de buenas prácticas de revisiones sistemáticas [29] y [30]. A diferencia de las revisiones sistemáticas, en los mapeos no es necesario determinar el rigor y la relevancia de los estudios primarios porque su objetivo es proporcionar una visión general del alcance del tema investigado [22]. Finalmente, los artículos seleccionados fueron importados y procesados con la herramienta de gestión de referencias bibliográficas Mendeley[1].

### 3.2. Extracción de datos

Se utilizó un cuestionario de extracción de datos basado en las preguntas de investigación para recopilar la información relevante de los estudios primarios. La extracción fue realizada por dos investigadores y revisada por un tercero, comprobando la información presente en el formulario con cada uno de los artículos para verificar la consistencia. Los datos recolectados incluyeron información general (título, autores, año de publicación y fuente) e información relativa a las preguntas de investigación (RQ1 – RQ5), como se ilustra en la Tabla III.

La información se extrajo exactamente como los autores la mencionan en los artículos y los conflictos se discutieron y resolvieron internamente por los investigadores. Se adoptó este enfoque para evitar la subjetividad y facilitar la replicación del estudio.

Para dar soporte a este proceso se utilizó la herramienta FRAMEndeley [31], una extensión que permite codificar con tablas el texto resaltado en el gestor de referencias Mendeley valiéndose de los colores de subrayado para distinguir y extraer fragmentos de texto relevantes. En este caso se le asignó un color identificativo a cada RQ.

Previo a la extracción, se realizó una prueba piloto con 15 artículos para calibrar el instrumento de extracción, refinar la estrategia y evitar diferencias entre los investigadores.

---

[1] https://www.mendeley.com

Tabla III: Formulario de extracción de datos.

| | ID | Item | Descripción |
|---|---|---|---|
| General | D1 | Título | Título del estudio primario |
| | D2 | Autores | Autores del estudio primario |
| | D3 | Año de Publicación | Año de publicación del estudio primario |
| | D4 | Fuente | Fuente donde se publicó el estudio primario |
| RQ1 | D5 | Criterio | Criterio de selección de los proyectos utilizados en el estudio primario |
| RQ2 | D6 | Lenguaje | Lenguaje de programación utilizado en los proyectos seleccionados |
| | D7 | Tipo de proyecto | Tipo de proyecto de acuerdo a su funcionalidad |
| RQ3 | D8 | Meta-datos | Meta-datos extraídos de los proyectos seleccionados en los estudios primarios |
| RQ4 | D9 | Herramientas | Herramientas de recolección de meta-datos |
| RQ5 | D10 | Análisis | Tipo de análisis estadísticos que se realiza con los meta-datos extraídos |

### 3.3. Análisis de los datos

Las respuestas a las RQ fueron analizadas de forma cualitativa mediante la codificación abierta de los textos encontrados en los estudios primarios recolectados [32]. La selección de esta estrategia se basó en la realizada por Janssen y Van der Voort en [33]. El primer y segundo autor realizaron el proceso de forma separada e identificaron los desacuerdos, que se resolvieron sumando un tercer investigador que visitaba nuevamente la fuente de información y tenía en cuenta las justificaciones dadas por los dos integrantes originales.

Como siguiente paso se utilizó la codificación cerrada [34] para identificar y resignificar las respuestas a las taxonomías encontradas. Nuevamente los dos primeros autores del estudio realizaron la codificación inicial y los desacuerdos fueron resueltos mediante la presentación a un tercer investigador. En la sección 4 de análisis y resultados se pueden encontrar las descripciones de las clasificaciones empleadas.

## 4. Análisis y resultados

En esta sección, se presentan los datos recopilados y se responde a las preguntas de investigación con los datos extraídos. Los mismos se organizaron y resumieron en tablas y gráficos para una mejor visualización. Para tener un mayor nivel de detalle, en el Anexo[2] están disponibles las planillas originales con los datos extraídos.

### 4.1. Selección de artículos

Se obtuvo un total de 122 estudios primarios aplicando la estrategia descrita en la Sección 3.1. La búsqueda se efectuó en la revista ESE y las conferencias EASE y ESEM empleando título, resumen y palabras claves indexadas. Los resultados se observan en la Tabla IV.

---

[2] http://bit.ly/AnexoTablaCompleta

Tabla IV: Resultados de búsqueda.

|      | EMSE | ESEM | EASE |
|------|------|------|------|
| 2013 | 37   | 65   | 31   |
| 2014 | 61   | 72   | 61   |
| 2015 | 55   | 41   | 30   |
| 2016 | 74   | 58   | 30   |
| 2017 | 91   | 70   | 50   |
| 2018 | 105  | 58   | 26   |
| 2019 | 119  | 54   | 43   |
| 2020 | 146  | 43   | 69   |
| Total| 688  | 461  | 340  |

En la búsqueda manual fueron recolectados 1489 artículos, de los cuales se incluyeron 1396 y excluyeron 93 según los criterios de exclusión. De los 1396 incluidos en la fase 2, 1285 fueron rechazados por no cumplir alguno de los criterios de inclusión. A los 111 artículos aceptados se le sumaron 11 en la fase 4, quedando finalmente la suma de 122. Los resultados se observan en la Tabla V.

Tabla V: Detalle de cada fase de selección de estudios primarios.

| Fase | Descripción | Incluidos | Excluidos |
|------|-------------|-----------|-----------|
| 1 | Búsqueda manual | 1489 | - |
| 2 | Selección por criterios de exclusión | 1396 | 93 |
| 3 | Selección por criterios de inclusión | 111 | 1285 |
| 4 | Bola de Nieve hacia atrás | 11 | - |

En la Fig. II se puede observar un gráfico de burbujas con la cantidad de estudios seleccionados por año en EMSE, ESEM y EASE. La revista EMSE aportó la mayor cantidad de artículos con un total de 69, seguido por ESEM con 32 y EASE con 10.

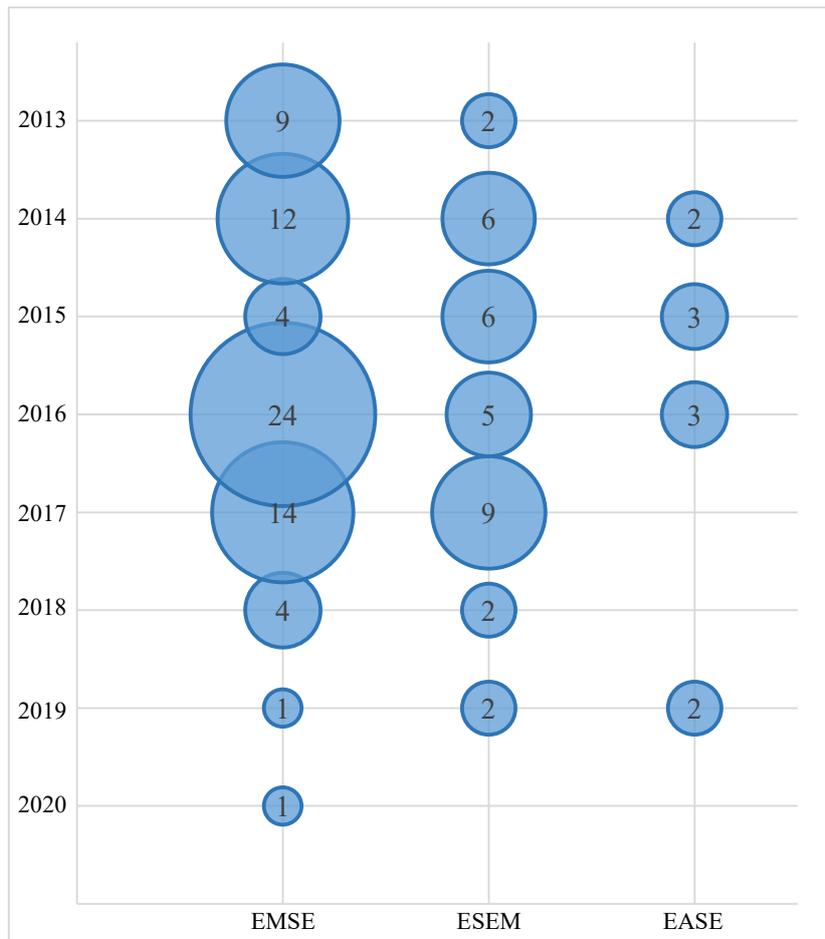

**Fig. II: Distribución de los artículos seleccionados de cada revista por año.**

Desde el año 2013 al 2018 se seleccionaron en promedio casi 20 artículos por año registrándose el punto máximo en el 2016 con 32. Desde el año 2018 en adelante este promedio se reduce a 4 artículos por año siendo el punto mínimo en el 2020 con un solo artículo.

### 4.2. RQ1: ¿Cuáles son los criterios de selección de los proyectos software objeto de estudios empíricos?

Los estudios recolectados fueron clasificados de acuerdo a la estrategia de selección escogida teniendo en cuenta dos enfoques, uno general y otro particular. En la primera clasificación (ver Fig. III), que abarca 3 categorías, el 72% de los artículos optaron por un método de selección propio, el 27% utilizaron una colección existente o subgrupo de esta y el 1% los recolectaron al azar.

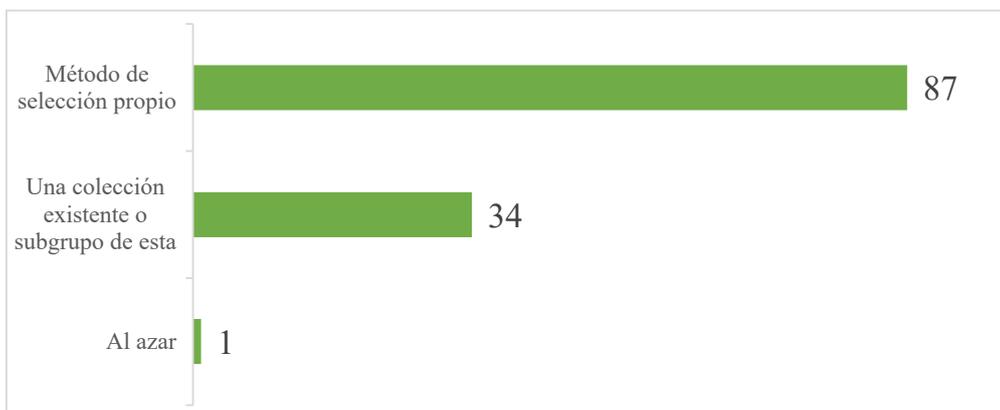

**Fig. III: Método utilizado para elegir los proyectos.**

En la Tabla VI se encuentran las colecciones que fueron utilizadas en los 33 estudios que implementaron esta estrategia. Algunos ejemplos son: los datasets de PROMISE, el SourceForge 100 (SF100), el Qualitas Corpus y el dataset del Grupo de Estándares y Benchmarking Internacional (ISBSG), entre otros.

**Tabla VI: Colecciones de proyectos.**

| Ocurrencias | Nombre de la colección | Artículos |
|---|---|---|
| 17 | PROMISE [35 - 45] | P3 P11 P15 P29 P57 P80 P96 P99 P106 P107 P112 P113 P116 P117 P119 P120 P122 |
| 2 | ISBSG [46] | P15 P86 |
| 2 | SF100 [P4] | P25 P60 |
| 2 | Qualitas Corpus [6] | P44 P121 |
| 1 | Tukutuku [47] | P3 |
| 1 | Centro de desarrollo de un banco de China | P18 |
| 1 | CVS-Vintage [48] | P21 |
| 1 | Sistemas de defensa de Corea del Sur | P23 |
| 1 | Mockus [49] | P27 |
| 1 | Mkaouer et al.[50] | P54 |
| 1 | Dataset de Finlandia | P58 |
| 1 | Hamasaki et al [51] | P59 |
| 1 | Departamento de defensa de Estados Unidos | P82 |
| 1 | Vasilescu, et al [52] | P98 |
| 1 | Yu et al. [53] | P103 |
| 1 | FlossMole [17] | P104 |
| 1 | Qualitas Corpus de aplicaciones en Python [54] | P105 |
| 1 | D'Ambros et al [55] | P112 |

Para la segunda clasificación (ver Tabla VII) se definieron 13 categorías. La clasificación fue elaborada en función de los datos recopilados tal y como se indica en el punto 3.3. En total, se extrajeron datos de 94 estudios primarios.

**Tabla VII: Criterios tenidos en cuenta para seleccionar los proyectos.**

| Ocurrencias | Criterios | Artículos |
|---|---|---|
| 34 | Específicas del caso de estudio del articulo | P8 P12 P13 P15 P23 P32 P34 P45 P48 P50 P51 P58 P61 P62 P64 P65 P66 P68 P69 P71 P73 P75 P76 P79 P82 P83 P89 P96 P98 P99 P102 P109 P110 P111 P114 |
| 32 | Proyectos y meta-datos disponibles públicamente | P1 P6 P8 P10 P14 P21 P26 P31 P35 P36 P39 P43 P46 P47 P50 P56 P64 P65 P68 P69 P72 P73 P74 P75 P76 P77 P78 P83 P94 P95 P98 P114 |
| 27 | Actividad en el proyecto a lo largo del tiempo | P1 P7 P9 P20 P27 P32 P34 P39 P42 P47 P49 P50 P54 P56 P59 P61 P68 P71 P77 P82 P86 P88 P90 P97 P111 P114 P115 P118 |
| 26 | Popularidad | P6 P7 P8 P20 P36 P37 P39 P43 P45 P56 P61 P65 P66 P68 P72 P73 P74 P77 P78 P79 P89 P91 P92 P100 P101 P115 |
| 24 | Dominio o tipo de software | P3 P7 P8 P9 P13 P14 P18 P22 P30 P34 P38 P42 P43 P44 P45 P55 P60 P70 P72 P92 P93 P100 P104 P115 |
| 24 | Tamaño | P6 P7 P11 P14 P15 P17 P34 P35 P42 P54 P58 P61 P65 P70 P72 P75 P88 P95 P99 P100 P102 P104 P114 P118 |
| 16 | Usan herramientas que dan soporte a procesos | P27 P51 P59 P63 P65 P67 P68 P71 P75 P88 P90 P92 P94 P97 P101 P102 |
| 12 | Actividad reciente en el momento de recolección | P1 P26 P27 P39 P54 P69 P74 P77 P78 P97 P98 P102 |
| 10 | Disponibilidad Información de defectos | P1 P46 P47 P50 P54 P73 P79 P94 P95 P111 |
| 8 | Equipo de Desarrollo | P6 P32 P35 P39 P56 P73 P114 P118 |
| 8 | Calidad | P6 P15 P19 P48 P58 P73 P76 P109 |
| 7 | Disponibilidad datos históricos | P1 P3 P7 P32 P56 P90 P115 |
| 5 | Documentación | P6 P7 P61 P73 P98 |

De esta manera, el 36% de los estudios describieron criterios específicos del caso de estudio por el cual se realizaba el experimento. Por ejemplo, repositorios cuyo primer commit no pareciera una migración en [P8], proyectos que incluyan una matriz que indique las fallas que cubren los tests en [P12], sistemas de fuente abierta que sean similares a soluciones industriales en [P13], entre otros.

El 34% de los estudios construyeron la colección de proyectos seleccionando aquellos que estuvieran disponibles públicamente con sus meta-datos asociados. Se mencionaron sitios de alojamiento de proyectos software como: SourceForge en [P1], [P6], [P26], [P50] y [P94]; ohloh.net en [P8]; Squeaksource en [P10]; Google Play en [P31] y [P43]; Github en [P36], [P39], [P65], [P68], [P72], [P74], [P75], [P76], [P77], [P83], [P98] y [P114] o Apple Store en [P43]; entre otros.

El 29% consideraron la actividad del proyecto a lo largo del tiempo como factor de selección. La actividad puede ser medida por cantidad de años de desarrollo o cantidad de commits realizados. Así se encuentran casos como en [P1] que selecciona proyectos con 3 o más años, en [P71] recolecta proyectos con más de 10K commits, o en [P50] que excluye aquellos sistemas que tengan menos de 32 commits y un año de desarrollo.

El 28% eligieron la popularidad en repositorios públicos como criterio de recolección. Dependiendo del trabajo se tomaron enfoques diferentes para cuantificarla. Por ejemplo, la popularidad se expresó términos de: el número de descargas [P6]; la cantidad de usuarios como [P7], [P20], [P45], [P56] y [P115]; el número de visitas [P36]; el número de duplicaciones del repositorio [P39]; o la cantidad de reseñas de usuarios [P43], entre otras.

En el 26% de los estudios primarios consideraron como criterio el dominio o tipo de software, es decir, la funcionalidad o ámbito de aplicación del mismo. En 21 artículos ([P3], [P7], [P8], [P13], [P14], [P22], [P30], [P34], [P38], [P42], [P43], [P44], [P55], [P60], [P70], [P72], [P92], [P93], [P100], [P104] y [P115]) mencionan que los sistemas seleccionados deben provenir de múltiples dominios de aplicación. En los tres casos restantes recolectaron proyectos de usos menos generales, como bases de datos en [P9], aplicaciones de propósito general en [P45], y ámbitos más específicos como sistemas de un banco comercial en China en [P18].

El 26% tuvieron en cuenta el tamaño de los proyectos cuantificado en términos de clases como en [P6], [P14], [P42] y [P65]; módulos en [P17]; líneas de código [P34], [P35], [P75], [P99] y [P118]; puntos de función IFPUG en [P15] o puntos de función FiSMA en [P58].

En el 17% de los casos una condición fue el uso de herramientas para dar soporte a procesos dentro del desarrollo del proyecto. Esto incluyó procesos tales como: gestión de dependencias en [P27], [P65], [P68], [P75], [P88], [P101] y [P102]; gestión de versiones en [P63], [P67], [P71], [P90], [P92], [P94] y [P97] o revisión de código en [P51] y [P59].

El 13% buscaron proyectos en los que se haya registrado actividad reciente al momento de recolección, es decir, que los mismos sigan siendo mantenidos o actualizados por sus colaboradores. La actividad suele ser medida en base a la cantidad de commits realizados en un periodo de tiempo [P98].

El 11% de los estudios establecieron como criterio de selección que hubiera disponibilidad de información de defectos en sistemas de rastreo de defectos.

En el 9% de los artículos recolectaron proyectos según la cantidad de participantes en el equipo de desarrollo o si existiera una comunidad que ofreciera soporte. También con un 9% fue considerada la calidad del proyecto, donde se utilizaron diversos métodos para establecerla. En [P48] tomaron proyectos orientados a objetos bien diseñados con evidencias de prácticas agiles; en [P73], [P76] y [P109] filtraron los proyectos de baja calidad con la herramienta Reaper; en [P15] y [P58] evaluaron la calidad según los métodos de IFPUG y FiSMA respectivamente.

El 7% buscaron aquellos proyectos en los que existiera datos históricos del proceso de desarrollo, en algunos casos para recuperar información evolutiva del sistema. Y en el 5% de los casos, la documentación del proyecto fue un factor determinante, en medios como los comentarios en el código fuente ([P6], [P61] y [P98]), documentación de desarrollo en [P7] y licencia del software en [P73].

### 4.3. RQ2: ¿Con qué tipo de proyectos trabajan los grupos de investigación para realizar estudios empíricos?

Cada uno de los proyectos software en las colecciones utilizados en los estudios seleccionados fueron clasificados según el lenguaje de programación principal (Fig. IV) y el tipo de software (Fig. V). En total se registraron 110 artículos que describieron los lenguajes de programación principales de los proyectos. En un 79% fueron construidos en Java, seguido por C con un 27% y C++ con un 24%.

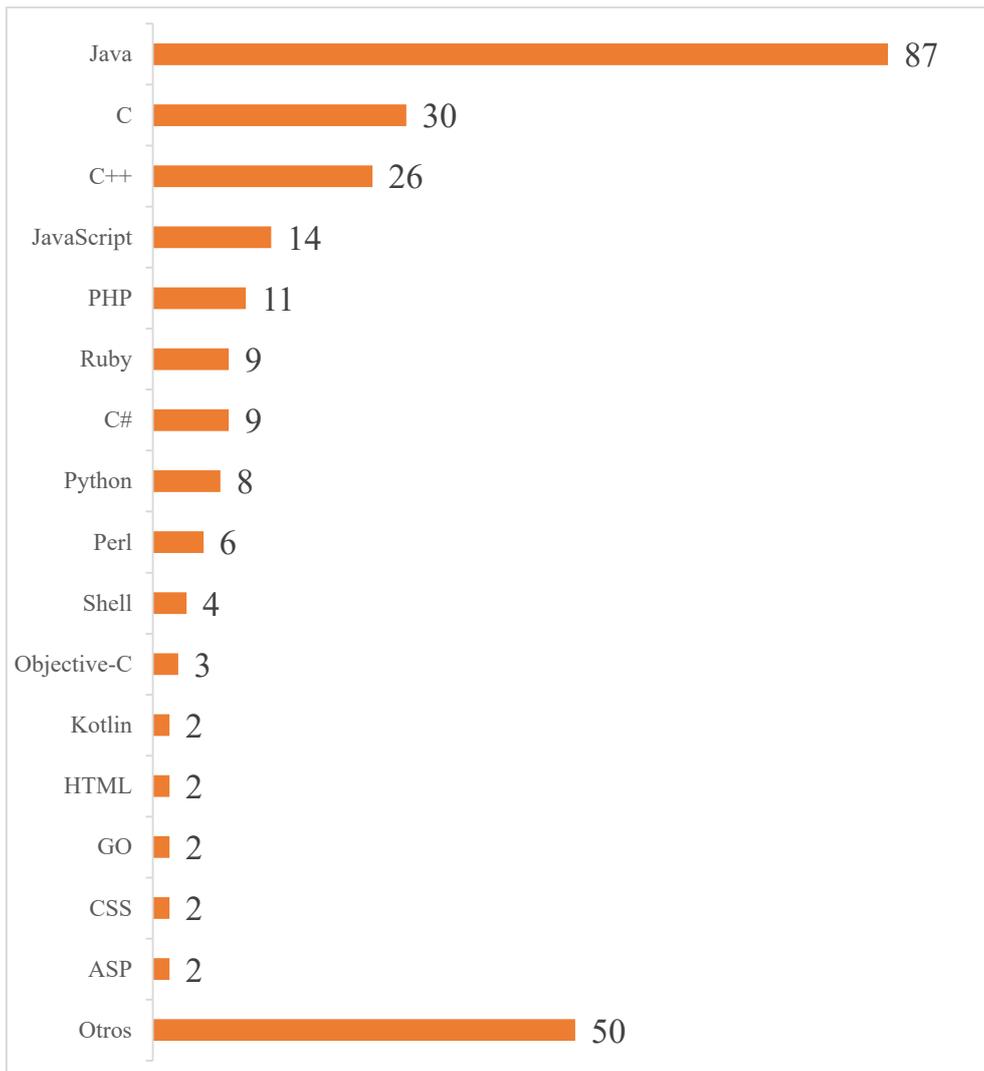

**Fig. IV: Lenguajes de programación de los proyectos.**

Para identificar los tipos de software se buscaron los nombres y descripciones de proyectos mencionadas en los estudios. En los casos donde los datos extraídos no eran suficientes para determinar el tipo de software utilizado, se realizó una búsqueda y consulta manuales. Cada proyecto fue clasificado según la taxonomía publicada en [56], de donde se emplearon cinco categorías de segundo nivel: diseño y construcción de software, servidor, redes y comunicaciones, sistema operativo y sistema embebido (ver Fig. V). Para comprender las demás categorías dentro de la taxonomía original que abarcan al software guiado por datos, orientado al consumidor y de propósito general se usó el término "aplicación de uso general". De los estudios primarios seleccionados 96 aportaron esta información.

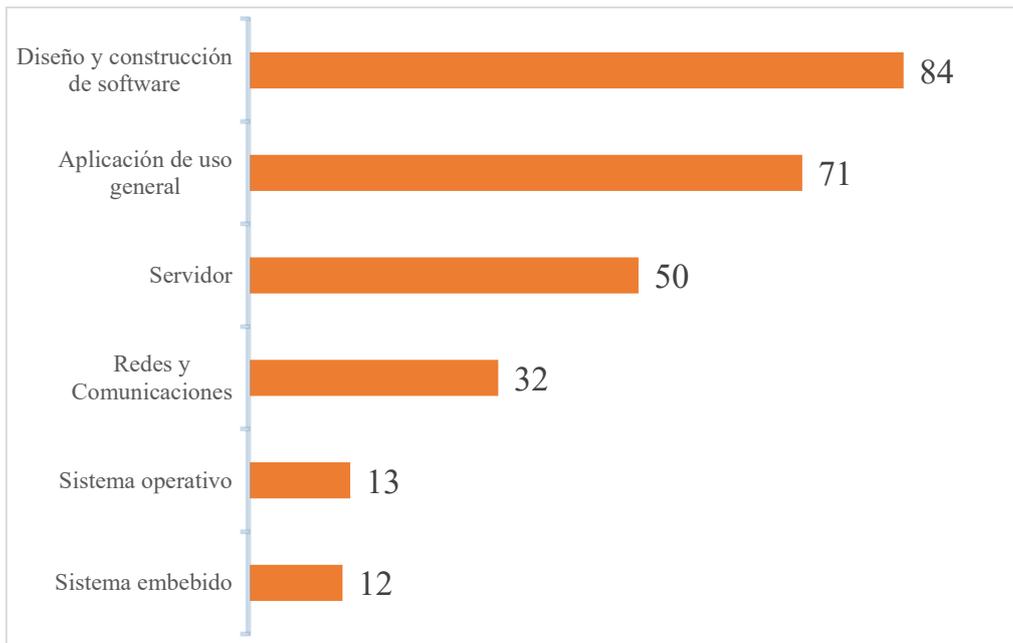

**Fig. V : Tipos de software.**

El 88% de los artículos trabajaron con proyectos software para el diseño y construcción de sistemas. En este sentido se consideran compiladores como: Clang en [P41], frameworks de desarrollo como Spring Framework en [P8], entornos de desarrollo integrado como Eclipse en [P2], interfaces de programación de aplicaciones como Apache iBatis en [P4], herramientas de construcción como Ant en [P8], herramientas para el modelado del sistema como ArgoUML en [P4] o librerías para la generación de casos de prueba como jUnit en [P5], entre otros.

El 74% experimentaron con aplicaciones de uso general. En esta clasificación entran: procesadores de texto como jEdit en [P1], videojuegos como Freecol en [P44], navegadores como Firefox en [P20], aplicaciones cliente de mensajería como Pidgin en [P1], sitios web como phpMyAdmin en [P1], aplicaciones android en [P64] o reproductores de música como aTunes en [P28], entre otras.

El 52% se identificó con servidores, es decir, sistemas preparados para la recepción de solicitudes de clientes. Por ejemplo, servidores web como Apache en [P4], de base de datos como MySQL en [P9], de transferencia de archivos como FileZilla en [P1], para computación distribuida como Hadoop en [P56] o de instancias de virtualización como Apache Cloudstack en [P71].

En el 33% utilizaron software de sistema para tareas de redes y comunicaciones entre aplicaciones y dispositivos, por ejemplo: Apache Synapse en [P11], dnsjava en [P21], Quagga en [P41], ActiveMQ en [P71]. El 14% trabajaron con sistemas operativos, tales como Linux en [P20] y Android en [P27].

En el 13% estudiaron sistemas embebidos, es decir, software embebido en un dispositivo mecánico o eléctrico diseñado para realizar una función específica. Por ejemplo, en [P23] utilizaron proyectos software armamentísticos de Corea del Sur, y en [P29] sistemas controladores de una compañía de Turquía. En [P34] sistemas industriales embebidos como controladores de motores de combustión o soluciones de procesamiento de audio.

### 4.4. RQ3: ¿Cuáles son los datos/meta-datos que se extraen de estos proyectos?

Dependiendo del objeto de estudio del artículo se recolectaron diferentes meta-datos del código de cada uno de los proyectos. En la Tabla VIII se agruparon los meta-datos siguiendo la taxonomía elaborada en [57] debido a la gran variedad existente. Las categorías consideradas de esta taxonomía son aquellas que abarcan métricas obtenibles por medio del análisis estático del código. De los 54 estudios que extrajeron meta-datos, las métricas presentes son de tamaño del código fuente en un 67%, por ejemplo, líneas de código (LOC) en [P14], volumen de Halstead en [P16] o líneas de comentarios también en [P16].

**Tabla VIII: Categorías de meta-datos.**

| Ocurrencias | Categorías de Meta-datos | Artículos |
|---|---|---|
| 36 | Tamaño Código Fuente | P8 P11 P14 P16 P23 P24 P26 P34 P36 P44 P50 P51 P53 P56 P67 P69 P71 P73 P74 P75 P76 P80 P82 P87 P92 P93 P96 P100 P107 P109 P112 P113 P114 P119 P120 P122 |
| 34 | Diseño | P5 P6 P11 P14 P16 P19 P26 P34 P35 P36 P44 P50 P51 P53 P54 P55 P65 P67 P74 P75 P76 P80 P93 P96 P97 P106 P107 P109 P112 P113 P116 P119 P121 P122 |
| 17 | Code Smells | P11 P31 P36 P37 P44 P54 P61 P66 P74 P75 P88 P90 P92 P100 P102 P109 P116 |

En el 63% de artículos se utilizan métricas de diseño que miden propiedades inherentes del software y se puede disgregar en otras 7 subcategorías: complejidad, acoplamiento, diagrama de clases, herencia, cohesión, encapsulamiento y polimorfismo (ver Tabla IX).

Las métricas de complejidad calculan la cantidad de caminos existentes en el flujo de control del programa; como la complejidad ciclomática de McCabe en [P26], o el peso de métodos por clase en [P11]. Las métricas de acoplamiento cuantifican la asociación o interdependencia entre los módulos, como el acoplamiento entre objetos en [P11] o la respuesta por clase en [P14].

Las medidas de diagrama de clases representan la estructura estática del sistema y pueden clasificarse en términos de clases, métodos y atributos (Tabla X), por ejemplo: el número de métodos en [P14], el número de clases en [P44] o el número de atributos en [P14].

Las métricas de herencia miden los atributos y métodos compartidos entre clases en una relación jerárquica, dentro de las cuales están el número de hijos en [P11], la profundidad de árbol de herencia en [P26]. Las métricas de cohesión establecen el grado en que métodos y atributos de una misma clase están conectados, como la falta de cohesión en los métodos, la cohesión de clases ajustada o la cohesión de clases suelta todas en [P14].

Las medidas de encapsulamiento calculan los datos y funciones empaquetadas en una sola unidad, por ejemplo, el número de métodos accessors en [P44], la métrica de acceso a datos en [P54]. Y las medidas de polimorfismo se usan para cuantificar la habilidad de un componente de tomar múltiples formas, como el número de métodos polimórficos en [P54].

**Tabla IX: Subcategorías de meta-datos de diseño.**

| Ocurrencias | Subcategorías de Diseño | Artículos |
|---|---|---|
| 23 | Complejidad | P11 P16 P26 P34 P36 P44 P50 P51 P67 P74 P75 P76 P80 P93 P97 P106 P107 P109 P112 P113 P116 P119 P122 |
| 19 | Acoplamiento | P5 P11 P14 P19 P26 P36 P44 P50 P53 P54 P67 P76 P80 P96 P106 P107 P116 P119 P122 |
| 18 | Diagrama de Clases | P14 P26 P35 P36 P44 P50 P53 P54 P65 P67 P76 P80 P93 P96 P106 P107 P119 P122 |
| 15 | Herencia | P6 P11 P26 P44 P50 P54 P67 P76 P80 P106 P107 P109 P116 P121 P122 |
| 15 | Cohesión | P11 P14 P26 P50 P54 P55 P67 P76 P80 P96 P106 P107 P116 P119 P122 |
| 9 | Encapsulamiento | P44 P53 P54 P67 P80 P96 P106 P107 P122 |
| 1 | Polimorfismo | P54 |

La tercera categoría es code smells. Los code smells son estructuras en el código fuente que a menudo indican la existencia de un problema de calidad o estructural, y plantean la necesidad de realizar una refactorización. Es una forma disciplinada de limpiar el código que reduce las chances de insertar defectos

[58]. Si bien esta categoría no se encuentra presente en la taxonomía original [57], es razonable incluirla en la misma porque se ha establecido como un método efectivo para descubrir problemas en el código fuente y por medio de la refactorización, mejorar la calidad y mantenimiento del software [P44]. Dicho esto, el 32% de los estudios recolectan code smells.

Tabla X: Subcategorías de meta-datos de diagrama de clases.

| Ocurrencias | Subcategorías de Diagrama de Clases | Artículos |
|---|---|---|
| 16 | Métodos | P14 P26 P35 P36 P44 P50 P54 P67 P76 P80 P93 P96 P106 P107 P119 P122 |
| 9 | Atributos | P14 P44 P54 P67 P80 P106 P107 P119 P122 |
| 5 | Clases | P44 P53 P54 P65 P76 |

### 4.5. RQ4: ¿Qué herramientas se utilizan para obtener estos datos/meta-datos?

Es común en estos estudios hallar diferentes herramientas para dar soporte a tareas específicas para lograr precisión y repetitividad [59]. Dada la variedad de herramientas encontradas, se realizó una clasificación en cinco categorías en base a las funciones que las mismas desempeñan. En la Fig. VI se puede observar un gráfico de burbujas con la clasificación antes mencionada (eje vertical), donde están descripta la cantidad de herramientas encontradas, cuántas de ellas siguen vigentes, es decir, si la última versión estable sigue en funcionamiento y cuantas han sido actualizadas después del 1º de enero del 2020 (eje horizontal). Dentro de las herramientas halladas se encuentran: Understand en [P26], CKJM en [P67] o PMD en [P11].

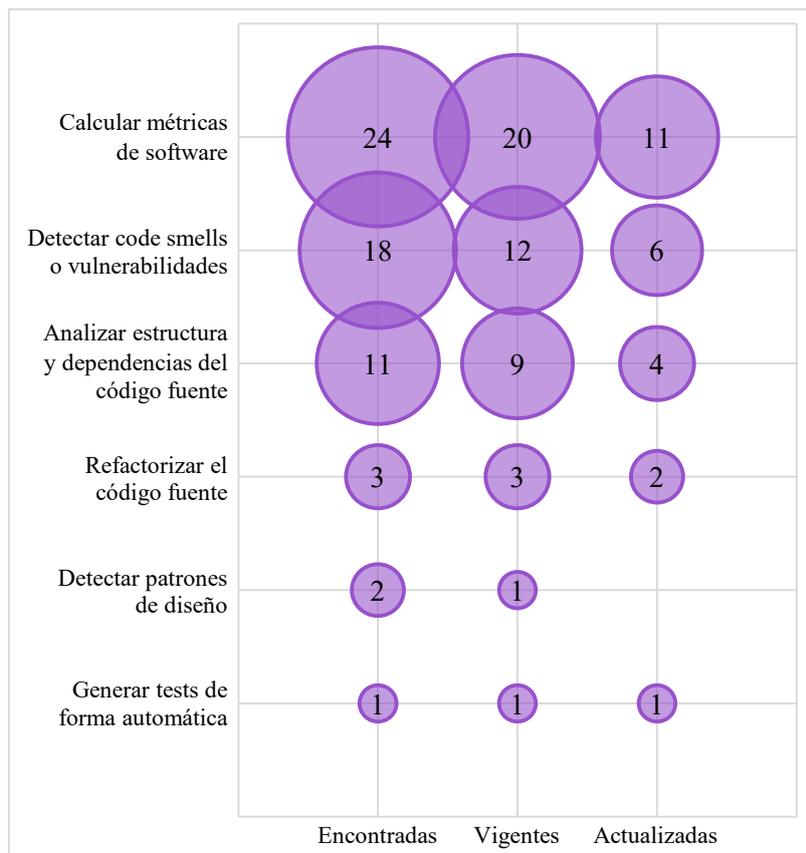

Fig. VI: Cantidad de herramientas encontradas, vigentes y actualizadas por tipo.

### 4.6. RQ5: ¿Qué análisis estadísticos se realizan con los datos/meta-datos recopilados?

Los investigadores necesariamente realizan un análisis previo a los datos recabados para exponer los hallazgos encontrados. Este proceso permite identificar las características que posee la muestra y a su vez seleccionar los estudios apropiados para generar los resultados. De los artículos recolectados se extrajeron los análisis estadísticos utilizados y se clasificaron en las dos áreas generales de la estadística: estadística inferencial y estadística descriptiva (ver Tabla XI), tomando como referencia las clasificaciones de [60]. En total 88 estudios informaron los procedimientos estadísticos realizados.

**Tabla XI. Clasificación de análisis estadísticos.**

| Ocurrencias | Estadística | Artículos |
|---|---|---|
| 69 | Inferencial | P1 P2 P3 P5 P6 P8 P10 P11 P12 P13 P15 P17 P19 P22 P25 P26 P27 P28 P29 P30 P32 P34 P35 P38 P42 P43 P46 P49 P50 P51 P53 P54 P56 P57 P58 P59 P60 P61 P62 P66 P67 P69 P70 P72 P73 P76 P77 P79 P80 P84 P85 P88 P89 P93 P95 P96 P100 P102 P103 P104 P105 P106 P107 P111 P113 P115 P116 P119 P122 |
| 64 | Descriptiva | P1 P2 P3 P4 P7 P8 P10 P12 P13 P14 P15 P17 P21 P25 P26 P28 P29 P30 P35 P36 P38 P42 P46 P47 P49 P50 P51 P53 P55 P56 P57 P58 P59 P60 P61 P66 P67 P68 P69 P70 P71 P73 P77 P78 P86 P88 P89 P92 P95 P98 P100 P102 P104 P106 P107 P108 P109 P111 P112 P114 P116 P117 P118 P122 |

La estadística inferencial emplea los datos para sacar conclusiones o hacer predicciones. En el 78% de los artículos seleccionaron procedimientos inferenciales que se clasificaron en 2 categorías, test no paramétrico y test paramétrico (ver Tabla XII).

Una de las diferencias entre estos grupos es que los test paramétricos hacen suposiciones específicas con respecto a uno o más parámetros de la población que caracterizan la distribución subyacente para la cual el test está siendo empleado. Ejemplos de test paramétricos son: test T en [P10], test chi cuadrado de Wald en [P51], y test de análisis de varianza (ANOVA) en [P1]. Dentro de los no paramétricos hay ejemplos como test Mann-Whitney en [P2], test de rangos con signo de Wilcoxon en [P3], y test de normalidad Shapiro-Wilk en [P28].

**Tabla XII. Procedimientos estadísticos inferenciales.**

| Ocurrencias | Procedimientos Inferenciales | | Artículos |
|---|---|---|---|
| 61 | Test Paramétrico | No | P1 P2 P3 P5 P6 P8 P10 P12 P13 P15 P17 P19 P22 P25 P26 P27 P28 P29 P30 P32 P34 P35 P38 P42 P43 P46 P49 P50 P51 P53 P54 P56 P57 P58 P59 P60 P61 P62 P66 P67 P69 P70 P72 P73 P76 P77 P79 P80 P85 P89 P93 P95 P100 P102 P104 P106 P107 P111 P115 P116 P122 |
| 22 | Test Paramétrico | | P1 P10 P11 P26 P27 P28 P46 P49 P51 P59 P60 P70 P76 P84 P88 P96 P103 P111 P113 P115 P119 P122 |

Por otra parte, el 73% de los artículos trabaja con procedimientos descriptivos utilizados para presentar y resumir los datos. Estos procedimientos fueron clasificados en cinco categorías: tamaño del efecto, medida de variabilidad, medida de tendencia central, medida de asimetría y medida de curtosis (ver Tabla XIII).

**Tabla XIII. Procedimientos estadísticos descriptivos.**

| Ocurrencias | Procedimientos Descriptivos | Artículos |
|---|---|---|
| 51 | Tamaño del efecto | P7 P8 P10 P12 P14 P17 P21 P25 P28 P29 P30 P35 P36 P38 P42 P47 P49 P50 P51 P53 P55 P56 P59 P60 P61 P66 P68 P69 P70 P71 P73 P77 P78 P86 P88 P89 P92 P95 P98 P100 P102 P104 P106 P107 P109 P111 P112 P116 P117 P118 P122 |
| 26 | Medida de Variabilidad | P1 P2 P3 P4 P10 P13 P14 P17 P26 P28 P30 P38 P46 P49 P51 P57 P58 P59 P67 P68 P88 P102 P108 P114 P116 P122 |
| 25 | Medida de Tendencia Central | P1 P2 P3 P4 P10 P13 P14 P17 P26 P28 P30 P38 P46 P49 P51 P58 P59 P67 P68 P88 P102 P108 P114 P116 P122 |
| 3 | Medida de Asimetría | P4 P15 P46 |
| 2 | Medida de Curtosis | P4 P46 |

El tamaño de efecto mide la magnitud de fuerza de un fenómeno o efecto. En Kitchenham et al. [61] remarcan su utilidad porque proporcionan una medida objetiva de la importancia que tiene un fenómeno en un experimento, independientemente de la significación estadística de la prueba de hipótesis conducida. Además, le afecta menos el tamaño de la muestra que a la significación estadística. Por ejemplo, podemos mencionar: el coeficiente de correlación Spearman en [P8], el tamaño de efecto de Cliff en [P28] o el coeficiente de correlación de Pearson en [P36].

Las medidas de variabilidad y las de tendencia central son los estadísticos más básicos empleados tanto en la estadística inferencial como en la descriptiva. En esta clasificación fueron incluidos solamente dentro de los procedimientos descriptivos porque en cada caso los usaron con el fin de describir la muestra recolectada. Como ejemplos de medidas de variabilidad se encuentran la desviación estándar en [P46], los cuartiles en [P14] y la varianza en [P108]. En el caso de medidas de tendencia central están la media en [P51], la mediana en [P4] y la moda en [P108].

## 5. Discusión

En esta sección se discuten los resultados obtenidos y como se relacionan con las preguntas de investigación identificadas en la sección 2. Sin embargo, es posible que no se hayan localizado todos los estudios relevantes, por esa razón el proceso fue desarrollado metodológicamente siguiendo un protocolo bien definido y múltiples investigadores revisaron la calidad de la información extraída. El mapeo comprendió 122 artículos publicados en la revista EMSE y las conferencias ESEM y EASE durante el período 2013 – 2020.

### 5.1. RQ1: ¿Cuáles son los criterios de selección de los proyectos software objeto de estudios empíricos?

Para abordar esta pregunta se buscaron evidencias de la selección de proyectos software con dos enfoques, uno a nivel general y otro más específico. A nivel general, en la mayoría de los casos (72%) los investigadores siguen lineamientos propios para seleccionar sus proyectos y en menor medida (27%) utilizan una colección de proyectos existentes. De los 94 estudios que reportaron información al respecto, los criterios de selección predominantes son: los específicos del caso de estudio (36%), junto con la disponibilidad de los proyectos y meta-datos (34%), el período de actividad en el proyecto (29%), la popularidad (28%), el dominio de aplicación (26%) el cual muchas veces no se encuentra descrito con precisión, el tamaño del proyecto (26%) que tiene en cuenta diferentes dimensiones, desde la cantidad de líneas de código hasta medidas de punto función.

La Tabla VII evidencia la diversidad de estrategias existentes para recolectar las muestras. En particular, algunos estudios optan por automatizar el proceso utilizando herramientas para explorar repositorios públicos como Github o SourceForge. Así, en [P69], [P76] y [P109] aplican un framework, que permite seleccionar repositorios Github que contengan "proyectos que aprovechan las prácticas sólidas de ingeniería de software en una o más de sus dimensiones, como la documentación, pruebas y gestión de proyectos" [P69]. O en [P94] y [P108] que utilizan un lenguaje de programación de dominio específico para el análisis y minado de repositorios software de gran escala [62].

En resumen, se evidencias estrategias diversas, pocos descritas o justificadas y en ocasiones sesgadas para los estudios.

### 5.2. RQ2: ¿Con qué tipo de proyectos trabajan los grupos de investigación para realizar estudios empíricos?

Para caracterizar la muestra de los proyectos software recolectados por los grupos de investigación se describieron el lenguaje de programación principal y el tipo de actividad o función que desempeña. De 110 artículos que reportan el lenguaje de programación principal de los proyectos, la mayoría (79%) experimentan con Java y en menor grado (31%) los lenguajes C o C++. Esto puede tener una relación directa con la cantidad de proyectos en los repositorios de ambos lenguajes [63]. Otra razón es la gran cantidad de herramientas de análisis estático del código fuente compatibles disponibles para el caso del lenguaje Java. Aún así, trabajos como [P3], [P23], [P27], [P39], [P49], [P50], [P69], [P73], [P114] y [P115] recolectaron proyectos con 5 o más lenguajes.

Finalmente, de los 96 artículos que fue posible determinar el tipo de software empleado la mayoría era para el diseño y construcción de sistemas (88%) y, en segundo lugar, aplicaciones de uso general (74%).

### 5.3. RQ3: ¿Cuáles son los datos/meta-datos que se extraen de estos proyectos?

Para responder esta pregunta se buscaron los meta-datos extraídos de los proyectos software para la realización de los experimentos. Así se encontraron 54 estudios primarios que reúnen meta-datos del código en forma de métricas, indicadores o medidas. De esta manera en este conjunto de artículos se pueden encontrar métricas que cuantifican el tamaño (67%), diseño (63%) y los code smells (32%). Esto evidencia el uso de meta-datos básicos en primera instancia y la posibilidad de incluir interpretaciones en una segunda instancia.

El resto de los trabajos se valieron de otras fuentes de información. Por ejemplo, en 13 artículos utilizaron los reportes de defectos para calcular su tiempo de resolución en [P1], [P46] y [P56]; clasificarlos en [P7], [P20], [P33] y [P52] o detectar reportes duplicados en [P47]; entre otros estudios.

En 6 artículos recolectaron la información de la ejecución del programa (análisis dinámico) para localizar funciones específicas en [P2], clasificar los procesos en [P18], estudiar la asignación de memoria en [P40], analizar cómo trabajan las excepciones en [P64] o las dependencias del sistema en ejecución en [P104].

En 50 artículos extrajeron los datos contenidos en el repositorio del proyecto para medir el acoplamiento y dependencias en la evolución del sistema en [P5], [P6] y [P32]; clasificar los commits en [P9] y [P24]; predecir defectos en el software en [P29]; estudiar el uso de patrones en la historia del proyecto en [P42] o el aporte de los desarrolladores [P51], entre otros. Para estos casos es necesario que los proyectos tengan un conjunto de meta-datos registrados a lo largo de su tiempo de desarrollo. Esto se consigue muchas veces en la gestión del proyecto con una herramienta de control de versiones y desarrollo colaborativo.

### 5.4. RQ4: ¿Qué herramientas se utilizan para obtener estos datos?

El objetivo de esta pregunta es conocer las herramientas utilizadas en la Ingeniería del Software para la recolección de meta-datos de proyectos. Es notable que solamente 61 artículos mencionan de manera explícita los nombres de las herramientas utilizadas. De los cuales 44 estudios informan herramientas que estrictamente generen meta-datos de los proyectos, siendo este un requisito indispensable para la replicabilidad de los experimentos. En muchas ocasiones se informa el modelo, técnica, procedimiento o algoritmo utilizado, pero no así la herramienta utilizada. Por ejemplo, en [P2] utilizan un modelo de fusión de datos compuesto por técnicas de recuperación de información, análisis dinámicos y minado Web para la localización de métodos que desarrollan funciones específicas del sistema. En [P3] implementan un algoritmo de búsqueda meta-heurístico en un modelo de aprendizaje automatizado para estimar el esfuerzo de desarrollo. En [P18] declararon el uso de un "método propio" para la refactorización de artefactos software. En [P14] trabajaron con una herramienta con la que calcular 29 métricas de código fuente pero no reportan su nombre, ni como acceder a la misma.

**Tabla XIV. Tipo de herramientas.**

| Ocurrencias | Tipo de Herramientas | Artículos |
|---|---|---|
| 25 | Cálculo métricas de software | P11 P16 P22 P23 P26 P35 P44 P50 P51 P54 P56 P63 P67 P69 P74 P75 P76 P88 P89 P93 P94 P102 P104 P107 P109 |
| 17 | Análisis de la estructura y dependencias del código fuente | P6 P16 P17 P26 P34 P42 P45 P50 P51 P68 P74 P76 P89 P93 P102 P104 P111 |
| 17 | Detección de code smells o vulnerabilidades | P11 P16 P22 P36 P37 P42 P44 P54 P66 P75 P88 P90 P100 P102 P107 P109 P116 |
| 4 | Refactorización de código | P55 P74 P76 P77 |
| 4 | Generación automática de tests | P4 P25 P60 P62 |
| 2 | Detección patrones de diseño | P42 P54 |

Dicho esto, en los 44 estudios (ver Tabla XIV) usaron herramientas para el cálculo de métricas de software (57%), análisis de la estructura y dependencias del código fuente (37%), detección de code smells o vulnerabilidades (37%), refactorización de código (9%), generación automática de tests (9%) y detección patrones de diseño (5%). En la Tabla XV se encuentran las cuatro herramientas más repetidas en los artículos seleccionados.

**Tabla XV: Herramientas más utilizadas.**

| Ocurrencias | Herramientas | Artículos |
|---|---|---|
| 8 | Understand[3] | P26 P50 P51 P74 P76 P89 P102 P104 |
| 4 | Evo Suite[4] | P4 P25 P60 P62 |
| 3 | CKJM[5] | P26 P67 P107 |
| 3 | PMD[6] | P11 P22 P44 |

Existen herramientas que se ubican en más de una categoría de la clasificación presentada en la Fig. VI y Tabla XIV. Understand, Alitheia Core y Analizo calculan métricas de software y analizan la estructura y dependencias del código fuente. PMD, iPlasma, SonarQube, Designite e InCode también calculan métricas y además detectan code smells y vulnerabilidades del código.

### 5.5. RQ5: ¿Qué análisis estadísticos se realizan con los datos recopilados?

Esta pregunta pretende determinar qué técnicas o procedimientos estadísticos son elegidos por los investigadores para validar los resultados de los experimentos realizados. En este sentido, la selección y aplicación apropiada de los métodos de análisis es una de las recomendaciones de Wohlin y Rainer en [64] para garantizar que la evidencia generada se presente correctamente y evitar que existan interpretaciones erróneas de los resultados.

De los 88 artículos que se registraron procedimientos estadísticos el 78% son inferenciales, de los cuales el 25% contiene pruebas o tests estadísticos paramétricos para el análisis de los datos recopilados. Esto implica supuestos, como que los datos obtenidos sigan una distribución normal, lo que podría no coincidir con la forma de selección de los proyectos o los mecanismos de extracción de los datos. Así, por ejemplo, muchos estudios utilizan reglas basadas en la disponibilidad del proyecto o su popularidad, pero no en la representatividad de los meta-datos con respecto a la población.

---

[3] https://www.scitools.com
[4] https://www.evosuite.org
[5] https://www.spinellis.gr/sw/ckjm
[6] https://pmd.github.io

Finalmente, en 51 estudios incorporaron un análisis de tamaño del efecto, siendo esta una recomendación indicada para los casos en que las muestras sean poco representativas y se utilicen técnicas paramétricas [61].

### 5.6. Amenazas a la validez

A continuación, se discuten las amenazas a la validez del estudio siguiendo el enfoque propuesto por [65]. Se consideraron cinco aspectos de validez.

**Validez de constructo.** La validez del constructo refleja hasta qué punto la metodología de la investigación representa a la estrategia del investigador y lo que se busca estudiar en las preguntas de investigación. Esta amenaza está presente al diseñar el instrumento de extracción de datos. La misma disminuyó implementando una prueba piloto de la tabla, tomando artículos al azar para completar una primera versión, y modificándola iterativamente según sea necesario hasta lograr la versión final.

**Validez interna.** Este aspecto de validez analiza los riesgos cuando se estudian relaciones causales [66]. Elegir los artículos y el período de tiempo adecuados son factores importantes que afectan la validez interna. La cantidad de artículos relacionados con la ISBE ha crecido mucho recientemente y las revistas de Ingeniería del Software tienen diferentes grados de aceptación para investigaciones empíricas. Para mitigar esta amenaza, se seleccionaron artículos de revistas y conferencias ampliamente aceptados en el ámbito de la Ingeniería del Software en términos de alcance y reputación. La subjetividad en la selección disminuyó mediante el proceso descrito en la Sección 3.1, realizando tantas iteraciones como fueron necesarias hasta obtener un grado de acuerdo alto.

**Validez externa.** La validez externa se refiere hasta qué punto es posible generalizar los resultados más allá del estudio. Los hallazgos aquí presentados se basan en las publicaciones pertenecientes a EMSE, EASE y ESEM. Si bien se desconoce si los resultados se pueden generalizar a artículos de otras revistas o conferencias, la investigación se basó en el análisis de 122 artículos y por tanto, puede considerarse representativa.

**Fiabilidad.** La fiabilidad evidencia hasta qué punto los resultados de la investigación son independientes de los investigadores. Es decir, si otro autor llevase a cabo el mismo estudio, los resultados deberían ser iguales o similares [65]. En este artículo, los métodos y procesos de investigación se describen en detalle para garantizar su reproducibilidad y en el anexo se incluyen las planillas originales con los datos extraídos.

**Sesgo de publicación.** El sesgo de publicación se refiere al "problema de que es más probable que se publiquen los resultados positivos de la investigación que los negativos" [67]. Este problema ocurre en cualquier revisión de literatura o estudio de mapeo. Sin embargo, en este caso, su efecto fue moderado porque nuestro estudio no pretende comparar resultados de investigación.

## 6. Conclusiones

En este mapeo sistemático se identificaron artículos en los que se realizaron estudios empíricos con colecciones de proyectos. Se abordaron cinco preguntas de investigación de estos estudios, como los criterios de selección de proyectos, su caracterización, los meta-datos recolectados en los estudios empíricos, las herramientas utilizadas para generar u obtener los meta-datos y los análisis estadísticos desarrollados con ellos. Por medio de una búsqueda manual inicial en la revista EMSE y las conferencias ESEM y EASE se obtuvieron 1496 artículos entre el 1 de enero del 2013 al 31 de diciembre del 2020. De los cuales 122 estudios fueron seleccionados después de aplicar los criterios de inclusión y exclusión definidos. A continuación, se presentan las respuestas a las preguntas de investigación.

Respecto de los criterios de selección del conjunto de proyectos, las practicas más comunes realizadas por los investigadores en este sentido es seguir lineamientos propios para seleccionar los proyectos y utilizar una colección de proyectos existente. No se ha evidenciado un marco unificado o automatizado para la selección de proyectos debido a la gran diversidad de aspectos considerados en los 94 estudios que reportaron criterios con un mayor nivel de detalle. En 35 estudios se informaron criterios específicos del caso de estudio y solo en 5 casos se reportaron el uso de herramientas para automatizar el proceso de selección en base a las reglas establecidas.

Con respecto a las características de los proyectos recolectados, el lenguaje de programación principal de los proyectos software con un gran margen de diferencia es Java (79%), el segundo y tercero son C y C++ (31%) de 110 artículos que se registraron respuestas. Los proyectos software utilizados fueron específicos con un 74% de aplicaciones de uso general y 96% de proyectos en el dominio del diseño y construcción de sistemas, servidores, redes, sistemas operativos o sistemas embebidos de los 96 artículos que fue posible determinar el tipo de software empleado.

Con respecto a los meta-datos de los proyectos, las fuentes de información de donde se extraen varían desde el mismo código fuente, la información en tiempo de ejecución, los reportes de defectos o el repositorio del proyecto, entre otras. En particular se encontraron 54 estudios primarios que reúnen meta-datos extraídos del código de los proyectos software en forma de métricas. Estas miden aspectos como el tamaño, diseño y code smells del proyecto.

Respecto de las herramientas para obtener los meta-datos, en 44 estudios primarios se utilizan herramientas para la recolección de meta-datos de proyectos obtenidos por medio del análisis estático del código. Realizan tareas como el cálculo de métricas de software, el análisis de la estructura y dependencias del código fuente, la detección de code smells o vulnerabilidades, la refactorización de código, la generación automática de tests y la detección patrones de diseño. Las herramientas que más se repitieron fueron Understand (8), Evo Suite (4), CKJM (3) y PMD (3).

De los análisis estadísticos desarrollados sobre los meta-datos, en 88 artículos se registraron tanto procedimientos estadísticos inferenciales como descriptivos. La mayoría de los métodos inferenciales fueron pruebas estadísticas no paramétricas (69%) y en menor medida paramétricas (25%). En el caso de los métodos descriptivos, se utilizaron tamaño de efecto, medidas de variabilidad y medidas de tendencia central. Dicho esto, en un gran número de casos no se observa que se tome en consideración la forma de selección de las muestras en los análisis estadísticos practicados, ignorando la posibilidad de que las mismas no sean representativas de la población.

Finalmente, como aporte de este trabajo se identificaron algunas pautas que ayudan a sistematizar la selección de proyectos en la construcción de colecciones con fines de investigación. Las principales reglas son: código fuente libre tanto su acceso como distribución, la vigencia del proyecto, su popularidad en repositorios y en lenguaje Java. La colección resultante debería conservar el código fuente de los proyectos, sus métricas obtenidas del análisis estático y los valores de estadísticos descriptivos que caractericen la muestra. Como trabajos futuros se analizarán las colecciones creadas con fines de investigación presentes en la Tabla VI y en artículos relacionados, con el objetivo de reconocer cuales fueron los criterios considerados en cada caso, el propósito de su construcción, y la vigencia de estos.

## Referencias

**Referencias estudios primarios**